\documentclass[twocolumn,aps,superscriptaddress,showpacs, nofootinbib,floatfix]{revtex4}
\usepackage{epsfig,bm}
\usepackage{graphics}
\usepackage{amsmath}
\usepackage{mathrsfs}
\begin{document}
\title{Proton-proton correlations in distinguishing the two-proton emission mechanism
of $^{23}$Al and $^{22}$Mg}
\author{D. Q. Fang}\thanks{dqfang@sinap.ac.cn}
\author{Y. G. Ma}\thanks{ygma@sinap.ac.cn}
\author{X. Y. Sun}
\affiliation{Shanghai Institute of Applied Physics, Chinese Academy of Sciences, Shanghai 201800, China}
\author{P. Zhou}
\affiliation{Shanghai Institute of Applied Physics, Chinese Academy of Sciences, Shanghai 201800, China}
\author{Y. Togano}
\affiliation{Institute of Physical and Chemical Research (RIKEN), Wako, Saitama 351-0198, Japan}
\author{N. Aoi}
\author{H. Baba}
\affiliation{Institute of Physical and Chemical Research (RIKEN), Wako, Saitama 351-0198, Japan}
\author{X. Z. Cai}\author{X. G. Cao}\author{J. G. Chen}\author{Y. Fu}\author{W. Guo}
\affiliation{Shanghai Institute of Applied Physics, Chinese Academy of Sciences, Shanghai 201800, China}
\author{Y. Hara}
\author{T. Honda}
\affiliation{Department of Physics, Rikkyo University, Tokyo 171-8501, Japan}
\author{Z. G. Hu}
\affiliation{Institute of Modern Physics, Chinese Academy of Sciences, Lanzhou 730000, China}
\author{K. Ieki}
\affiliation{Department of Physics, Rikkyo University, Tokyo 171-8501, Japan}
\author{Y. Ishibashi}
\author{Y. Ito}
\affiliation{Institute of Physics, University of Tsukuba, Ibaraki 305-8571, Japan}
\author{N. Iwasa}
\affiliation{Department of Physics, Tohoku University, Miyagi 980-8578, Japan}
\author{S. Kanno}
\affiliation{Institute of Physical and Chemical Research (RIKEN), Wako, Saitama 351-0198, Japan}
\author{T. Kawabata}
\affiliation{Center for Nuclear Study (CNS), University of Tokyo, Saitama 351-0198, Japan}
\author{H. Kimura}
\affiliation{Department of Physics, University of Tokyo, Tokyo 113-0033, Japan}
\author{Y. Kondo}
\affiliation{Institute of Physical and Chemical Research (RIKEN), Wako, Saitama 351-0198, Japan}
\author{K. Kurita}
\affiliation{Department of Physics, Rikkyo University, Tokyo 171-8501, Japan}
\author{M. Kurokawa}
\affiliation{Institute of Physical and Chemical Research (RIKEN), Wako, Saitama 351-0198, Japan}
\author{T. Moriguchi}
\affiliation{Institute of Physics, University of Tsukuba, Ibaraki 305-8571, Japan}
\author{H. Murakami}
\affiliation{Institute of Physical and Chemical Research (RIKEN), Wako, Saitama 351-0198, Japan}
\author{H. Ooishi}
\affiliation{Institute of Physics, University of Tsukuba, Ibaraki 305-8571, Japan}
\author{K. Okada}
\affiliation{Department of Physics, Rikkyo University, Tokyo 171-8501, Japan}
\author{S. Ota}
\affiliation{Center for Nuclear Study (CNS), University of Tokyo, Saitama 351-0198, Japan}
\author{A. Ozawa}
\affiliation{Institute of Physics, University of Tsukuba, Ibaraki 305-8571, Japan}
\author{H. Sakurai}
\affiliation{Institute of Physical and Chemical Research (RIKEN), Wako, Saitama 351-0198, Japan}
\author{S. Shimoura}
\affiliation{Center for Nuclear Study (CNS), University of Tokyo, Saitama 351-0198, Japan}
\author{R. Shioda}
\affiliation{Department of Physics, Rikkyo University, Tokyo 171-8501, Japan}
\author{E. Takeshita}\author{S. Takeuchi}
\affiliation{Institute of Physical and Chemical Research (RIKEN), Wako, Saitama 351-0198, Japan}
\author{W. D. Tian}\author{H. W. Wang}
\affiliation{Shanghai Institute of Applied Physics, Chinese Academy of Sciences, Shanghai 201800, China}
\author{J. S. Wang}\author{M. Wang}
\affiliation{Institute of Modern Physics, Chinese Academy of Sciences, Lanzhou 730000, China}
\author{K. Yamada}
\affiliation{Institute of Physical and Chemical Research (RIKEN), Wako, Saitama 351-0198, Japan}
\author{Y. Yamada}
\affiliation{Department of Physics, Rikkyo University, Tokyo 171-8501, Japan}
\author{Y. Yasuda}
\affiliation{Institute of Physics, University of Tsukuba, Ibaraki 305-8571, Japan}
\author{K. Yoneda}
\affiliation{Institute of Physical and Chemical Research (RIKEN), Wako, Saitama 351-0198, Japan}
\author{G. Q. Zhang}
\affiliation{Shanghai Institute of Applied Physics, Chinese Academy of Sciences, Shanghai 201800, China}
\author{T. Motobayashi}
\affiliation{Institute of Physical and Chemical Research (RIKEN), Wako, Saitama 351-0198, Japan}

\date{\today}
\begin{abstract}
The proton-proton momentum correlation functions ($C_{pp}(q)$) for kinematically complete decay 
channels of $^{23}$Al $\rightarrow$ p + p + $^{21}$Na and $^{22}$Mg $\rightarrow$ p + p + $^{20}$Ne 
have been measured at the RIKEN RI Beam Factory. From the very different correlation strength of 
$C_{pp}(q)$ for $^{23}$Al and $^{22}$Mg, the source size and emission time information were extracted from 
the $C_{pp}(q)$ data by assuming a Gaussian source profile in the correlation function calculation code (CRAB).
The results indicated that the mechanism of two-proton emission from $^{23}$Al was mainly
sequential emission, while that of $^{22}$Mg was mainly three-body simultaneous emission.
By combining our earlier results of the two-proton relative momentum and the opening angle,
it is pointed out that the mechanism of two-proton emission could be distinguished clearly. 
\end{abstract}
\maketitle

\section{Introduction}
The two-particle intensity interferometry has been extensively utilized to determine the
space-time extension of particle emitting sources in nuclear and particle physics over the past several
decades~\cite{hbt,goldhar,Koonin,Lynch,Bertulani,Led,Lisa,Hu,Verde,ma06}. In heavy ion collisions,
the two-particle interferometry is a well-recognized  and powerful method to characterize the source 
of particle emission and to probe and disentangle different reaction mechanisms. 
In particular, this method can provide information on the space-time evolution of hot nuclei 
which usually decay by evaporation and/or (multi-)fragmentation. Even though a large number 
of experiments have been carried out to measure the two-proton correlation in nuclear fragmentation, 
almost no proton-proton momentum correlation function measurement was reported for a kinematically 
complete decay channel. In contrast, there have been several measurements of the neutron-neutron 
momentum correlation function in kinematically complete decay channels for halo nuclei, 
such as $^{11}$Li~\cite{ieki,orr,Marq} and $^{14}$Be~\cite{orr}, which were believed to be  useful  
for studying the so-called di-neutron structure of neutron-rich nuclei~\cite{neut}.

The phenomenon of two-proton emission is a very interesting but complicated process existing in
the nucleus close to the proton-drip line~\cite{gold60,pfutzner,blank08,Olsen,brown}. The proton-proton 
correlation plays an important role in the emission mechanism. There is a distinct difference in the spectra 
of the two-proton relative momentum ($q_{pp}$) and opening angle ($\theta_{pp}$) between 
the diproton emission and two-body sequential or three-body simultaneous emission. 
This characteristic has been used to investigate the mechanism of two-proton emissions~\cite{kryger,raci}.

The proton-rich nuclei $^{23}$Al and $^{22}$Mg are very important in determining some astrophysical 
reaction rates and have attracted a lot of attention in both astrophysics and nuclear structure 
studies~\cite{iacob,gade,wies88,cai,fang,ozawa,wies99,sewe}.
Recently, we have reported the experimental results for kinematically complete measurements of 
two-proton emissions from two excited  proton-rich nuclei, namely
$^{23}$Al $\rightarrow$ p + p + $^{21}$Na and 
$^{22}$Mg $\rightarrow$ p + p + $^{20}$Ne~\cite{ma2015}.
Based on the analysis of $q_{pp}$ and $\theta_{pp}$ distributions of the two emitted protons, 
a favorable diproton emission component from the excited states around 14.044 MeV of $^{22}$Mg 
was observed. But no such signal exhibited in the two-proton emission processes of 
the excited $^{23}$Al. 

As pointed out in Ref.~\cite{ma2015}, it is difficult to distinguish between
the two-body sequential and three-body simultaneous emission mechanism using the above analysis. 
In these two mechanisms, the emission time of the two protons is quite different. 
For three-body simultaneous emission, 
the two protons are emitted almost at the same time, while the two protons are emitted one by one in  
sequential emission. The two-particle intensity interferometry method has been demonstrated to 
be a good way to extract the source size and particle emission time~\cite{lisa93,verde06}.
In this paper, the proton-proton momentum correlation function will be studied for 
the three-body decay channels $^{23}$Al $\rightarrow$ p + p + $^{21}$Na and 
$^{22}$Mg $\rightarrow$ p + p + $^{20}$Ne. The size and emission time information of the source 
will be extracted. The possibility of distinguishing sequential and three-body simultaneous emission 
mechanism will be investigated.

\section{Experiment Description}
\begin{figure}[h]
\begin{center}\includegraphics[width=7.5cm]{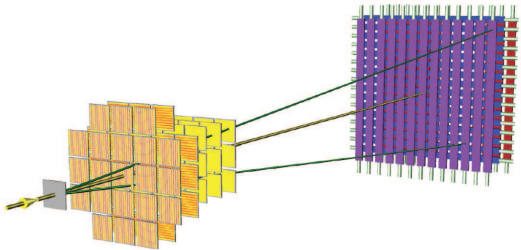}\end{center} 
\vspace{-0.4cm}
\caption{(Color online) The layout of detector setup. For details see the text.}
\label{setup}
\end{figure}

The experiment was performed using the RIPS beamline at the RI Beam Factory (RIBF) operated by 
RIKEN Nishina Center and Center for Nuclear Study, University of Tokyo. 
A primary beam of $135A$ MeV $^{28}$Si was used to produce secondary $^{23}$Al
and $^{22}$Mg beams with incident energy of 57.4$A$ MeV and 53.5$A$ MeV in the center of 
the carbon reaction target, respectively. After the reaction target, there were five layers of silicon 
detectors and three layers of plastic hodoscopes as shown in Fig.~\ref{setup}.
The first two layers of Si-strip detectors located around 50 cm downstream of the target were 
used to measure the emitting angle of the fragment and protons. Three layers of  3$\times$3 single-electrode 
Si were used as the $\Delta E$-$E$ detectors for the fragment. The three layers of plastic hodoscopes 
located around 3 m downstream of the target were used as $\Delta E$ and $E$ detectors for protons. 
TOF of proton was measured by the first layer. Clear particle identifications were obtained by this setup 
for the kinematically complete three-body decay reactions. The momentum and emission angle for 
protons and the residue are determined by analyzing the detector signals. The excitation energy ($E^*$) 
of the incident nucleus was reconstructed by the difference between the invariant mass of three-body 
system and mass of the mother nucleus in the ground state. More detailed description of 
the experiment could be found in Ref.~\cite{ma2015}.

\section{Results and discussion}
In the present study, the momentum correlation functions between two protons emitted from 
$^{23}$Al and $^{22}$Mg are studied. Experimentally, the two-proton correlation function is constructed
through dividing the coincidence yield $N_c$ by the yield of non-correlated events $N_{nc}$, namely
$C_{pp}(q) = K\frac{N_c(q)}{N_{nc}(q)}$,
where the relative momentum is given by $q=\frac{1}{2}|{\bf p}_1-{\bf p}_2|$, 
with ${\bf p}_1$ and ${\bf p}_2$ being the momenta of the two coincident protons. 
The normalization constant $K$ is determined so that the correlation function
goes to unity at large values of $q$, where no correlation is expected.

\begin{figure}[t]
\begin{center}\includegraphics[width=7.5cm]{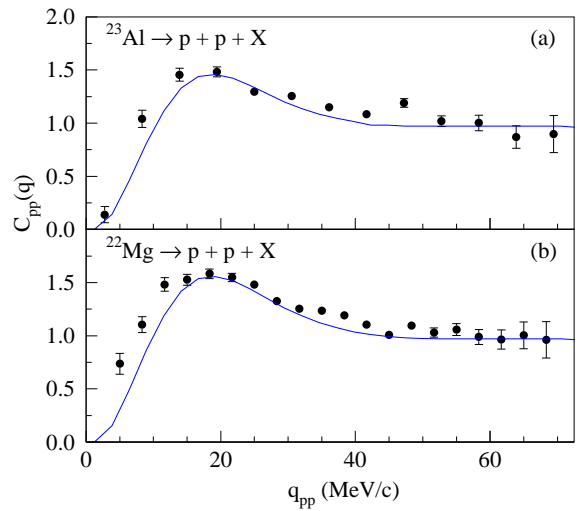}\end{center}
\vspace{-0.5cm}
\caption{(Color online) The proton-proton momentum correlation function ($C_{pp}(q)$) for the reaction 
channel of $^{23}$Al $\rightarrow$ p + p + X (a) and $^{22}$Mg $\rightarrow$ p + p + X (b).
The dots are experimental data. The lines are the calculations by the CRAB code with
a Gaussian source.}
\label{cppx}
\end{figure}

The event-mixing technique~\cite{hbt} was applied to  construct the background yield,
i.e. by pairing  a proton  with a randomly chosen uncorrelated proton
from different events and then normalized to the number of two-proton correlation events in
$N_c$. This method ensures that the uncorrelated distribution includes the same class of
collisions and kinematical constraints as the numerator. It has, however, a potential problem:
it may attenuate the slight correlations one wishes to measure~\cite{goldhar}
due to the existence of possible  "residue correlation" from initial two-proton physical correlation,
which will overestimate  the  denominator. To eliminate this "residue correlation", an iterative 
calculation method for the $C_{pp}(q)$ was applied and intrinsic correlation was extracted~\cite{Zajc}.
A similar method has been first applied to two-neutron momentum correlation function
measurement for neutron-halo nuclei~\cite{Marq}.
Here it is the first attempt to apply the correlation function analysis on two-proton emission data.

Firstly, we looked at the two-proton correlation in the inclusive reaction channel which is similar to the 
proton-proton momentum correlation function for hot nuclei. Fig.~\ref{cppx} showed our measurements 
of $C_{pp}(q)$ which were obtained by the event-mixing method with an iterative calculation for 
the two emitted protons, without identifying the residue from the mother nucleus and using 
any specific $E^*$ window. Fig.~\ref{cppx}(a) and Fig.~\ref{cppx}(b) were the results for the inclusive 
channel of $^{23}$Al $\rightarrow$ p + p + X and $^{22}$Mg $\rightarrow$ p + p + X, respectively. 
In this work, the normalization factor $K$ in calculating $C_{pp}(q)$ was determined by 
the $50<q_{pp}<100$ MeV/c data.
The peak height around $q_{pp}=20$ MeV/c reflected the strength of correlation function. 
To extract the source size, theoretical calculation for $C_{pp}(q)$ was performed by using 
the  correlation function calculation code (CRAB)~\cite{pratt}. 
A Gaussian profile was assumed for the source and the space distribution 
was simulated according to the function $S(r) \sim \exp(-r^2/2r_0^2)$, with $r_0$ being the source 
size parameter. The calculations were compared with the experimental $C_{pp}(q)$ data. 
The source size was determined by finding the minimum of the reduced chi-square ($\chi^2/\nu$, 
$\nu$ is the degree of freedom). The fit gave a source size range of $r_0=3.15\sim3.25$ fm for 
$^{23}$Al (corresponding to the rms radius of $R_{rms}=5.46\sim5.63$ fm). The uncertainty was 
determined from the minimum chi-square $\chi_0^2$ to $\chi_0^2+1$. The best fit of the calculation 
was plotted in the figure. The obtained source size could give us information of the average 
distance between the two emitted protons. This size is much larger than the expected radius of the 
$^{23}$Al nucleus. Of course, a caution needs to be taken for the value of $r_0$, which should be 
considered as the apparent size for the source since the emission time between two protons is another 
folded ingredient. For $^{22}$Mg, the source size was extracted to be $r_0=2.9\sim3.0$ fm which is 
a little bit smaller than that of $^{23}$Al.

\begin{figure}[t]
\begin{center}\includegraphics[width=7.5cm]{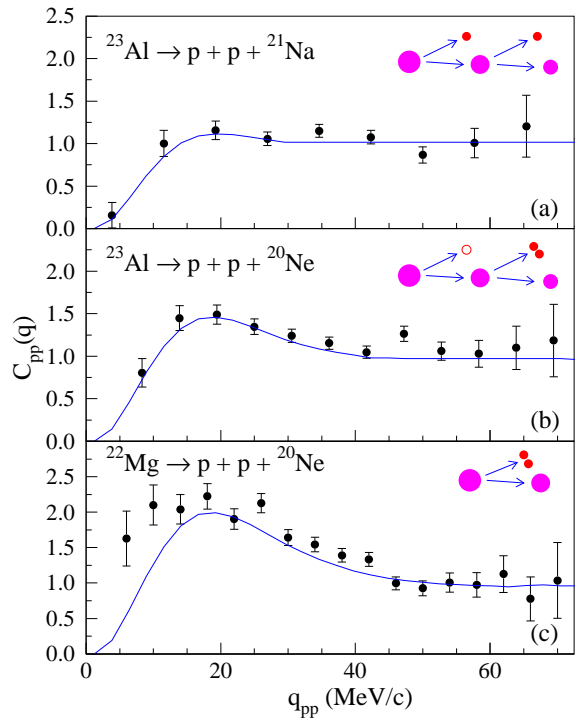}\end{center}
\vspace{-0.5cm}
\caption{(Color online) Same as Fig.~\ref{cppx} but for the reaction channel of $^{23}$Al $\rightarrow$ 
p + p + $^{21}$Na (a); $^{23}$Al $\rightarrow$ p + p + $^{20}$Ne (b); and $^{22}$Mg $\rightarrow$ 
p + p + $^{20}$Ne (c). Inserts show sketch maps for the most probable emission mechanism.}
\label{cpp}
\end{figure}

Secondly, we checked the kinematically complete three-body channels for both $^{23}$Al and $^{22}$Mg.
Fig.~\ref{cpp} showed the $C_{pp}(q)$ of the two emitted protons which were coincident with the
residues from the mother nuclei. For the channel of $^{23}$Al $\rightarrow$ p + p + $^{21}$Na, 
Fig.~\ref{cpp}(a) showed an almost flat correlation function except for the Coulomb dip 
in low $q_{pp}$ region, indicating that emission of both protons from $^{23}$Al three-body break-up 
was uncorrelated in phase space except for the Coulomb interaction. This was consistent with the result 
of no clear observation of diproton emission from the relative momentum and opening angle spectra of 
$^{23}$Al at any excited states~\cite{ma2015}.  Generally speaking, a flat proton-proton momentum 
correlation function indicates a very large source size and very weak two-proton correlation. 
The $C_{pp}(q)$ data was also compared with 
the Gaussian source calculations. The fit gave a source size $r_0=3.9\sim4.7$ fm which was larger than 
that of the inclusive channel of $^{23}$Al, indicating a more loose two-proton emission. 
Since the effect of emission time was not considered, it is difficult to explain the results only by 
the geometric size of the source.

In contrast, the $C_{pp}$ data for $^{22}$Mg nucleus was very different from that of $^{23}$Al
as shown  in Fig.~\ref{cpp}(c). In this figure, a strong correlation emerged in $C_{pp}$ spectra for 
the process of $^{22}$Mg $\rightarrow$ p + p + $^{20}$Ne, which indicated a compact source size of 
two-proton emission. The fit gave $r_0=2.35\sim2.45$ fm, which was smaller than that of the inclusive channel 
of $^{22}$Mg. 

Between the two very different two-proton correlation pattern of $^{23}$Al and $^{22}$Mg, we had checked the 
intermediate situation. If we looked at the decay process of $^{23}$Al $\rightarrow$ p + p + $^{20}$Ne,
where one proton was not detected by the experimental setup, a moderate correlation appeared
at $q_{pp}\sim20$ MeV/c as shown in Fig.~\ref{cpp}(b), which could be understood by assuming the
following two-step proton emission process of $^{23}$Al. One proton was emitted from $^{23}$Al 
and its corresponding residue nucleus was $^{22}$Mg; Then the  other two protons were ejected from 
$^{22}$Mg and its corresponding residue nucleus was $^{20}$Ne.
Among the three emitted protons, only two protons were detected by the detectors.
Because of a strong two-proton correlation in the second step, a moderate two-proton correlation could be 
eventually observed in the process of $^{23}$Al $\rightarrow$ p + p + $^{20}$Ne.
The peak height of $C_{pp}$ in Fig.~\ref{cpp}(b) could be explained by a mixture of
Fig.~\ref{cpp}(a) and Fig.~\ref{cpp}(c). The Gaussian source fit gave a size of $r_0=3.1\sim3.3$ fm, 
which was between the cases of Fig.~\ref{cpp}(a) and Fig.~\ref{cpp}(c).  
In addition, to give a visual impression of two-proton emission, the sketch maps were plotted as inserts 
in Fig.~\ref{cpp} to illustrate the most probable emission mechanism for each channel. 

\begin{figure}[t]
\begin{center}\includegraphics[width=7.2cm]{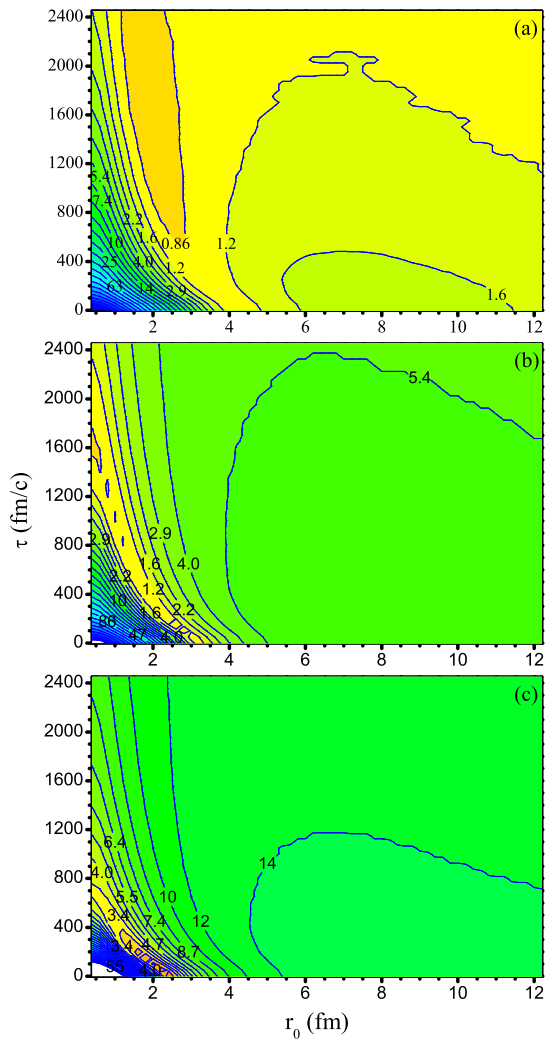}\end{center}
\vspace{-0.5cm}
\caption{(Color online) Contour plot of the reduced chi-square ($\chi^2/\nu$) obtained 
from fitting the proton-proton momentum correlation function by the CRAB calculation for 
the reaction channel of $^{23}$Al $\rightarrow$ p + p + $^{21}$Na (a); 
$^{23}$Al $\rightarrow$ p + p + $^{20}$Ne (b); and $^{22}$Mg $\rightarrow$ p + p + $^{20}$Ne (c).} 
\label{chi3}
\end{figure}

The effective source size was extracted from the above analysis. However, it was not sure how the 
two protons were emitted, i.e. the two protons were emitted sequentially or simultaneously. 
In these two cases, opposite values of the effective source size were observed for $^{23}$Al and $^{22}$Mg. 
It indicated that the emission time of protons might be different for these two nuclei. Since time information could also 
be extracted from the correlation function, it will be very interesting to extract both source size and 
emission time information. Thus a more general analysis was done for the experimental $C_{pp}(q)$ data.
For the different mechanism of two-proton emission, the emission time difference between the two protons is 
important. Assuming the first proton being emitted at time $t=0$ and the second proton being emitted 
at time $t$, the space and time profile of the Gaussian source was simulated according to the function 
$S(r,t) \sim \exp(-r^2/2r_0^2-t/\tau)$ in the CRAB code. $\tau$ refers to the lifetime for the emission 
of the second proton, which starts from the emission time of the first proton. The agreement between the 
calculation and the $C_{pp}(q)$ data was evaluated by determining the value of the reduced chi-square. 
The results were shown in Fig.~\ref{chi3} by a contour plot of $\chi^2/\nu$ as a function of 
$r_0$ and $\tau$. For the reaction channel of $^{23}$Al $\rightarrow$ p + p + $^{21}$Na as shown 
in Fig.~\ref{chi3}(a), the ranges of source parameters were obtain to be $r_0=1.2\sim2.8$ fm and 
$\tau=600\sim2450$ fm/c based on the best chi-square fit. While for the reaction channel of 
$^{22}$Mg $\rightarrow$ p + p + $^{20}$Ne as shown in Fig.~\ref{chi3}(c), the ranges of source parameters 
were $r_0=2.2\sim2.4$ fm and $\tau=0\sim50$ fm/c.
It means that the emission time difference between two protons for $^{23}$Al and $^{22}$Mg was 
quite different. For $^{23}$Al, the two protons were emitted at very different time ($\tau>600$ fm/c), 
i.e. the mechanism is a sequential emission. For $^{22}$Mg, the two protons were 
emitted almost at the same time ($\tau<50$ fm/c), i.e. the mechanism was essentially simultaneous. 
Based on the above results and the $q_{pp}$ and $\theta_{pp}$ analysis in Ref.~\cite{ma2015}, 
all observables indicate the three-body simultaneous decay mechanism for $^{22}$Mg. 
Moreover, for the excitations around the $14.044$ MeV, T = 2 state a strong diproton-like component was observed.
For the reaction channel of $^{23}$Al $\rightarrow$ p + p + $^{20}$Ne 
as shown in Fig.~\ref{chi3}(c), the determined source parameters were $r_0=0.4\sim2.0$ fm and 
$\tau=350\sim1950$ fm/c, which were also between the above two cases and could be explained 
by the two-step proton emission process of $^{23}$Al.

\section{Conclusions}
In summary, measurement on the proton-proton momentum correlation function was applied to kinematically 
complete decay of two reaction channels $^{23}$Al $\rightarrow$ p + p + $^{21}$Na and 
$^{22}$Mg $\rightarrow$ p + p + $^{20}$Ne in this paper. The experiment was performed at the RIKEN RI Beam Factory.
The proton-proton momentum correlation function $C_{pp}$ was obtained by the event-mixing method 
with an iterative calculation. By assuming a simple Gaussian emission source, the effective source sizes 
were extracted by comparing the CRAB calculation with the experimental $C_{pp}$ data. 
Different effective source size was obtained for $^{23}$Al and $^{22}$Mg, which comes from 
the different mechanism of two-proton emission. In a more general analysis including source size and 
emission time information, a reasonable source size but completely different emission time for 
the two protons were extracted. The results indicated that the mechanism of two-proton emission 
from $^{23}$Al was dominately sequential, while that for $^{22}$Mg was mainly three-body 
simultaneous emission with a strong diproton-like component at excited states around $14.044$ MeV. 
Based on the previous results~\cite{ma2015} and this work, it is possible to distinguish clearly
the mechanism of two-proton emission by investigating on the proton-proton momentum 
correlation function, the two-proton relative momentum and opening angle distributions.
The method presented in this work was applied for the first time to two-proton emitters, 
and was shown to provide new and valuable information on the mechanism of two-proton emission.

\section*{Acknowledgements}
We are very grateful to all of the staffs at the RIKEN accelerator for providing beams during 
the experiment. The Chinese collaborators greatly appreciate the hospitality from the RIKEN-RIBS 
laboratory. This work is supported by the Major State Basic Research Development Program of 
China under contract No. 2013CB834405, National Natural Science Foundation of China under 
contract Nos. 11421505, 11475244 and 11175231.


\begin{thebibliography}{0}
\bibitem{hbt} R. Hanbry Brown and R. Q. Twiss,  Nature {\bf178},  1046 (1956).
\bibitem{goldhar}G. Goldhaber, S. Goldhaber, W. Lee,  A. Pais,  Phys. Rev.  {\bf 120},  300 (1960).
\bibitem{Koonin} S. E. Koonin, Phys. Lett. {\bf 70B}, 43 (1977).
\bibitem{Lynch}W. G. Lynch {\it et al.}, Phys. Rev. Lett. {\bf 51}, 1850 (1983).
\bibitem{Verde} G. Verde {\it et al.}, Phys. Rev. C {\bf 65}, 054609 (2002).
\bibitem{Bertulani}C. A. Bertulani, M. S. Hussein, G. Verde, Phys. Lett.  B {\bf 666}, 86 (2008).
\bibitem{Led} R. Lednicky, V.L. Lyuboshitz, Yad. Fiz.  {\bf 35}, 1316 (1982) (Sov. J. Nucl. Phys. {\bf 35}, 770 ( 1982)).
\bibitem{Lisa}M. Lisa, S. Pratt, R. Soltz, U. Wiedemann, Ann. Rev. Nucl. Part. Sci. {\bf  55},  357 (2005).
\bibitem{Hu}Y. Hu, Z. Su, W. Zhang,	Nucl. Sci. Tech. {\bf 24}, 050522 (2013). 
\bibitem{ma06}Y. G. Ma {\it  et al.}, Phys. Rev. C {\bf 73}, 014604 (2006); Nucl. Phys. A {\bf 790}, 299c (2007).
\bibitem{ieki}K. Ieki {\it et al.}, Phys. Rev. Lett. {\bf 70}, 730 (1993).
\bibitem{orr}N. A. Orr,  Nucl. Phys. A {\bf  616},  155 (1997).
\bibitem{Marq}F. M. Marques {\it et al.}, Phys. Lett. B {\bf 476}, 219 (2000); Phys. Rev. C {\bf  64}, 061301(R) (2001).
\bibitem{neut}Z. Kohley {\it et al.}, Phys. Rev. Lett. {\bf 110}, 152501 (2013); 
                       E. Lunderberg {\it et al.}, Phys. Rev. Lett. {\bf 108}, 142503 (2012).
\bibitem{gold60} V. I. Goldansky, Nucl. Phys. {\bf 19}, 482 (1960).
\bibitem{pfutzner}M. Pfutzner, M. Karny,  L. V. Grigorenko and K. Riisager, Rev. Mod. Phys. {\bf84},  567 (2012), 
                              and references therein.
\bibitem{blank08}B. Blank and M. Ploszajczak, Rep. Prog. Phys. {\bf71},  046301 (2008), and references therein.
\bibitem{Olsen}E. Olsen {\it et al.}, Phys. Rev. Lett. {\bf 110}, 222501 (2013).
\bibitem{brown}K. W. Brown {\it et al.}, Phys. Rev. Lett. {\bf 113}, 232501 (2014).
\bibitem{kryger}R. A. Kryger {\it et al.}, Phys. Rev. Lett.  {\bf 74},  860 (1995).
\bibitem{raci}G. Raciti {\it et al.}, Phys. Rev. Lett.  {\bf 100},  192503 (2008).
\bibitem{iacob} V. E. Iacob  {\it et al.},  Phys. Rev. C {\bf 74},  045810 (2006).
\bibitem{gade}A. Gade {\it et al.},  Phys. Lett. B {\bf 666},  218 (2008).
\bibitem{wies88}M. Wiescher {\it et al.}, Nucl. Phys. A{\bf 484}, 90 (1988); 
                          J. A. Caggiano {\it et al.}, Phys. Rev. C {\bf 64}, 025802 (2001).
\bibitem{cai}X. Z. Cai {\it et al.}, Phys. Rev. C {\bf 65},  024610 (2002).
\bibitem{fang} D. Q. Fang {\it et al.}, Phys. Rev. C {\bf 76},  031601(R) (2007).
\bibitem{ozawa}A. Ozawa {\it et al.}, Phys. Rev. C {\bf 74}, 021301(R) (2006).
\bibitem{wies99} M. Wiescher {\it et al.}, J. Phys. G {\bf 25}, R133 (1999).
\bibitem{sewe} D. Seweryniak {\it et al.}, Phys. Rev. Lett. {\bf 94}, 032501 (2005).
\bibitem{ma2015}Y. G. Ma, D. Q. Fang {\it et al.}, Phys. Lett.  B {\bf 743}, 306 (2015).
\bibitem{lisa93}M. Lisa {\it et al.}, Phys. Rev. Lett. {\bf 71}, 2863 (1993).
\bibitem{verde06}G. Verde, A. Chbihi, R. Ghetti, J. Helgesson, Eur. Phys. J. A {\bf 30}, 81 (2006).
\bibitem{Zajc}W. A. Zajc {\it et al.}, Phys. Rev. C {\bf 29}, 2173 (1984).
\bibitem{pratt}S. Pratt {\it et al.}, Nucl. Phys. A {\bf 566}, 103c (1994).
\end{thebibliography}
\end{document}